A. Korenev, G. Belokrylov, B. Lodonova, A. Novokhrestov

# Attacks on the neural network and defense methods

Abstract: This article will discuss the use of attacks on a neural network trained on audio data, as well as possible methods of protection against these attacks. FGSM, PGD and CW attacks, as well as data poisoning, will be considered. Within the framework of protection, Art-IBM and advertorch libraries will be considered. The obtained accuracy metrics within the framework of attack applications are presented

The aim of this paper is to consider the application of FGSM, PGD, CW and data poisoning attacks within the framework of using a neural network trained on audio data.

**Dataset**

The data set chosen for training the model is AudioMNIST [1], which consists of 30,000 audio files in wav format. These files contain voice recordings of 60 speakers who pronounce numbers from 0 to 9. There are 500 recordings for each speaker, respectively. The task is to determine the spoken word (number).

**Neural network**

The architecture is a convolutional neural network. Within each convolutional layer, the convolution parameters themselves are set (the size of the input and output space, the kernel size, stride, padding), the RELU activation function, and normalization using BatchNorm2d). After the convolutional layers, there is an adaptive averaging pooling layer and a fully connected layer. Before feeding the neural network with data, the following transformations are performed:

– Augmentation in the form of a shift of the tensor obtained from the signal. Also, a tensor filled with zeros is added to the beginning of the signal;

– The resulting augmented signal is converted into a Mel-spectrogram with parameters such as sampling frequency of 48000 Hz, number of points for generating fast Fourier transform equal to 2034, lengths of transitions between short-term Fourier transform windows, and filters in the amount of 64. The maximum amplitude value is also set to 80 dB.

Then the obtained data set from Mel-spectrograms was divided into training, validation and test sets. The training set was 0.8 of the original amount of data. The validation and test sets were 0.12 and 0.08, respectively. The neural network is faced with the task of classifying the obtained Mel-spectrograms into a certain class, which is a number from 0 to 9. Table 1 presents the neural network architecture in detail..

Table 1
**Neural network architecture**

| Layer | Input | Output | Kernel size | Kernel Shift | Indent |
|---|---|---|---|---|---|
| 1st convolutional layer | 1 | 8 | (5, 5) | (2, 2) | (2, 2) |
| 2nd convolutional layer | 8 | 16 | (3, 3) | (2, 2) | (1, 1) |
| 3rd convolutional layer | 16 | 32 | (3, 3) | (2, 2) | (1, 1) |
| 4th convolutional layer | 32 | 64 | (3, 3) | (2, 2) | (1, 1) |
| Pooling layer | Dimensionality of the output space = 1 ||||||
| Fully connected layer | Input space = 64 |||| Output space = 10 ||

**Neural network training**

The following parameters were set for training the neural network::

– Loss function - cross entropy;

– Adam Optimizer with Learning Rate 0.001;

– Number of epochs – 5.

The accuracy for the training dataset was 0.99.

The accuracy for the validation dataset was 0.99.

The accuracy for the test dataset was 0.99.

**Attacks**

The FGSM algorithm calculates the direction in the input space along which, from the image location, movement is the closest path to misclassification. This direction is calculated using gradient descent and a loss function. Once the direction is determined, the FGSM method introduces a tiny distortion into each input value, adding it if the direction of the malicious change is positive and subtracting it otherwise.

For FGSM [2], the Eps parameter was set (to introduce distortion into the data) and L-inf is the norm. The initial value of Eps is 0.05, in the cycle it was increased by 0.05 and re-applied against the model. The upper limit was set to 1. The lowest accuracy value (0.17) on the test data set using FGSM was achieved with eps equal to 0.95.

Next, the PGD attack was considered. This method is an iterative version of the FGSM attack. Thus, it is essentially an iterative introduction of minimal distortions of the image going beyond the class boundary, which becomes malicious. Also, the gradient direction is checked at each iteration in case the model is not linear. Instead of one step (as in FGSM) of size eps in the direction of the gradient sign, several smaller alpha steps are made, and the result is cut off by the same step. The following parameters were used:

– Eps = 0.2 (maximum distortion of the unfavorable example compared to the original input signal);

– eps_iter = 0.1 (step size for each attack iteration);

– nb_iter = 5 (number of iterations).

Similarly, the initial value of Eps is 0.05, incremented by 0.05 in the loop and re-applied against the model. The upper bound of eps is 1. The lowest accuracy value (0) on the test dataset using PGD was achieved with an eps of 0.95.

Next, the CW attack method [3] was considered. The CW method iteratively minimizes the L-norm, which is used to estimate the magnitude of change, ensuring the maximum difference in confidence between the target class and the most probable one closest to it. This increases the threshold above which a malicious image ceases to be malicious. The following parameters were set:

– n_classes=10 (number of target marks);

– lr=0.01 (learning rate);

– max_iterations=200 (Maximum number of iterations).

When considering various values of the parameters of this method, no noticeable changes in the drop in accuracy were found.

The accuracy on the test set after applying the attack was 0.36.

The last method was data poisoning. [4-5].

The original signal was noised as follows:

– The tensor was generated randomly from a uniform distribution.

– This tensor was then added to the original signal.

– The noisy signal was sent to the training and testing dataset.

The accuracy after applying data poisoning on the test dataset was 0.17.

It is worth noting that one separate signal from the original data set was considered as an experiment in pure and poisoned form. The differences were in the form of spectrograms, but the signal was indistinguishable by ear. In this case, the signal was the pronunciation of the word "One". Attempts were also made to recognize the signal in its original form and in the noisy one. The SPT systems recognized this signal as "One" in both cases.

**Defense**

Attempts were also made to restore the neural network after the attack, or to use protective measures before the attacks were applied.

One of the libraries that contains such methods is Art-IBM. However, as it turned out, this library was tailored for neural network architectures based on Tensorflow or Keras, which ultimately led to errors in the work of protective methods, or did not help restore the accuracy rate close to the original.

For neural networks built on Torch, there is a library called advertorch for conducting various attacks and protecting against them. However, the methods used there are aimed at datasets containing images (MNIST and CIFAR), which most likely led to the failure of their application in this case, since the resulting data is primarily audio data, for which the methods of existing libraries were not developed in this case.

**Conclusion**

Thus, various attacks on neural networks such as FGSM, PGD, CW, and data poisoning attack were considered. Satisfactory results from the attacker's point of view were obtained.

The capabilities of protection against attacks carried out using ready-made libraries have been tested, however, as a result of their application, low accuracy of the model or incompatibility in its current configuration have been obtained. This is bad, since there is a suspicion that it will be necessary to use manually written attack methods, or change the framework on which the neural network architecture is built, which in turn will lead to a new cycle of training and attacks.

*Acknowledgments*

This research was funded by the Ministry of Science and Higher Education of Russia, Government Order for 2023–2025, project no. FEWM-2023-0015 (TUSUR).

___________________________________